\documentclass[notitlepage,
 amsmath,amssymb,
 aps,
]{revtex4-1}

\usepackage{graphicx,epstopdf}
\usepackage{bm}

\begin{document}


\title{Spatio-temporal Characterization of Thermal Fluctuations in a Non-turbulent Rayleigh-B{\'e}nard Convection at Steady State}

\author{Yash Yadati}
\author{Nicholas Mears}
\author{Atanu Chatterjee}
\address{Department of Physics, Worcester Polytechnic Institute, 100 Institute Road, Worcester, MA, USA, 01605}

\date{\today}

\begin{abstract}
In this paper we present a detailed description of the statistical and computational techniques that were employed to study a driven far-from-equilibrium steady-state Rayleigh-B{\'e}nard system in the non-turbulent regime ($Ra\leq 3500$). In our previous work on the Rayleigh-B{\'e}nard convection system we try to answer two key open problems that are of great interest in contemporary physics: (i) how does an out-of-equilibrium steady-state differ from an equilibrium state and (ii) how do we explain the spontaneous emergence of stable structures and simultaneously interpret the physical notion of temperature when out-of-equilibrium. We believe that this paper will offer a useful repository of the technical details for a first principles study of similar kind. In addition, we are also hopeful that our work will spur considerable interest in the community which will lead to the development of more sophisticated and novel techniques to study far-from-equilibrium behavior.
\end{abstract}

\maketitle


\section{Introduction} 
The Rayleigh-B{\'e}nard convection system holds a place of special interest in the scientific community~\cite{bodenschatz2000recent,cross1993pattern,behringer1985rayleigh}. It is one of the oldest and most widely used canonical examples to study pattern formation and emergent behavior~\cite{cross1993pattern,koschmieder1993benard,jaeger2010far,heylighen2001science}. The wide variety of conceptual richness offered by the Rayleigh-B{\'e}nard convection makes it an active area of research in the thermal, fluids and complexity science disciplines~\cite{bodenschatz2000recent,cross1993pattern,koschmieder1993benard,zhang1993deterministic,shishkina2010boundary,kadanoff2001turbulent}. It is infact one of the simplest complex systems' that can be easily recreated in a laboratory with minimal efforts. Although, the current state of the art experimental setups, data logging techniques, numerical and mechanistic simulations have provided numerous critical insights about the fluid mechanical aspects, a lot of the thermodynamical interpretations still remain unresolved~\cite{niemela2000turbulent,chilla2012new,du2000turbulent,du2001temperature,pandey2018turbulent,stevens2011prandtl}. The simple idea that a thin film of viscous liquid when heated from the bottom at steady-state, gives rise to a plethora of emergent patterns that are stationary in time is inviting enough to invoke deep theoretical curiosities about the far-from-equilibrium interpretation of the (thermodynamic) equation of state~\cite{lucia2008probability,garcia2008thermodynamics,gallavotti2019nonequilibrium,chatterjee2016energy,chatterjee2016thermodynamics,yadati2018detailed,lieb1998guide,martyushev2006maximum,chatterjee2013principle,georgiev2016road}. In a recent series of papers, we have outlined these open problems through a systematic thermodynamic study of a non-turbulent Rayleigh-B{\'e}nard convection system at steady-state~\cite{chatterjee2018many,chatterjee2018non,2018arXiv181206002C,srinivasarao2019biologically}. This paper serves as a repository of experimental details, computational and analytical techniques. We strongly believe that the ideas laid out in this paper will not only serve as an evidence of reproducibility but will also spur considerable interest in the search for more efficient and novel techniques in the study of far-from-equilibrium thermodynamics in the future.

Over the years the Rayleigh-B{\'e}nard system has seen many state-of-the-art developments which have successfully helped us understanding the fluid mechanical aspects of the system along with theoretical and computational insights in turbulence studies~\cite{shishkina2010boundary,du2000turbulent,du2001temperature,pandey2018turbulent,stevens2011prandtl,du1998enhanced,stevens2010radial,schumacher2008lagrangian,grossmann2004fluctuations}. However, from a first principles thermodynamic point of view this system has a lot of insights to offer which have largely gone unnoticed. In recent series of papers we have investigated as how the thermodynamics of a far-from-equilibrium steady-state differs that from an equilibrium state. We have tried to quantify and characterize the differences by studying out-from-equilibrium thermal fluctuations in a driven Rayleigh-B{\'e}nard system. In this paper, we present the computational and statistical details of our results with the hope that this will serve as a reference for all future works in this line of thinking~\cite{chatterjee2018many,2018arXiv181206002C,vilar2001thermodynamics,lucia2013stationary}.

\section{Experimental Methodology}
\begin{figure}[]
\centering
\includegraphics[scale=0.65]{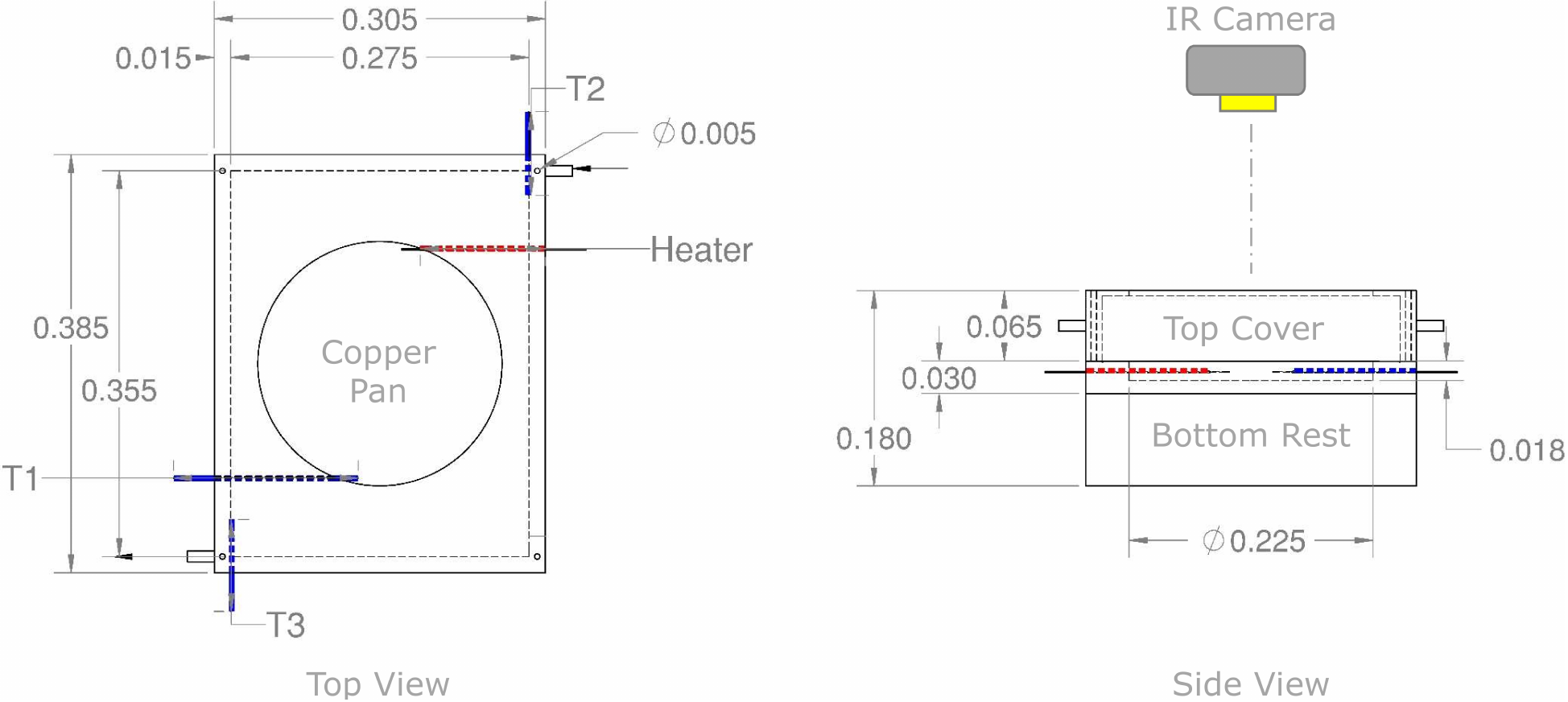}
\caption{Figure shows Rayleigh-B{\'e}nard Convection setup: Top View (left) and Side View (right). All dimensions are in meters. The three thermocouples used to record the temperature of the pan, the inlet and the outlet of the heat-exchanger are denoted in blue (T1, T2 and T3). The heater attached to the bottom of the copper pan is denoted in red. The thermal images are recorded from the top using an IR camera.}
\label{setup}
\end{figure}

\begin{figure}[]
\centering
\includegraphics[scale=0.5]{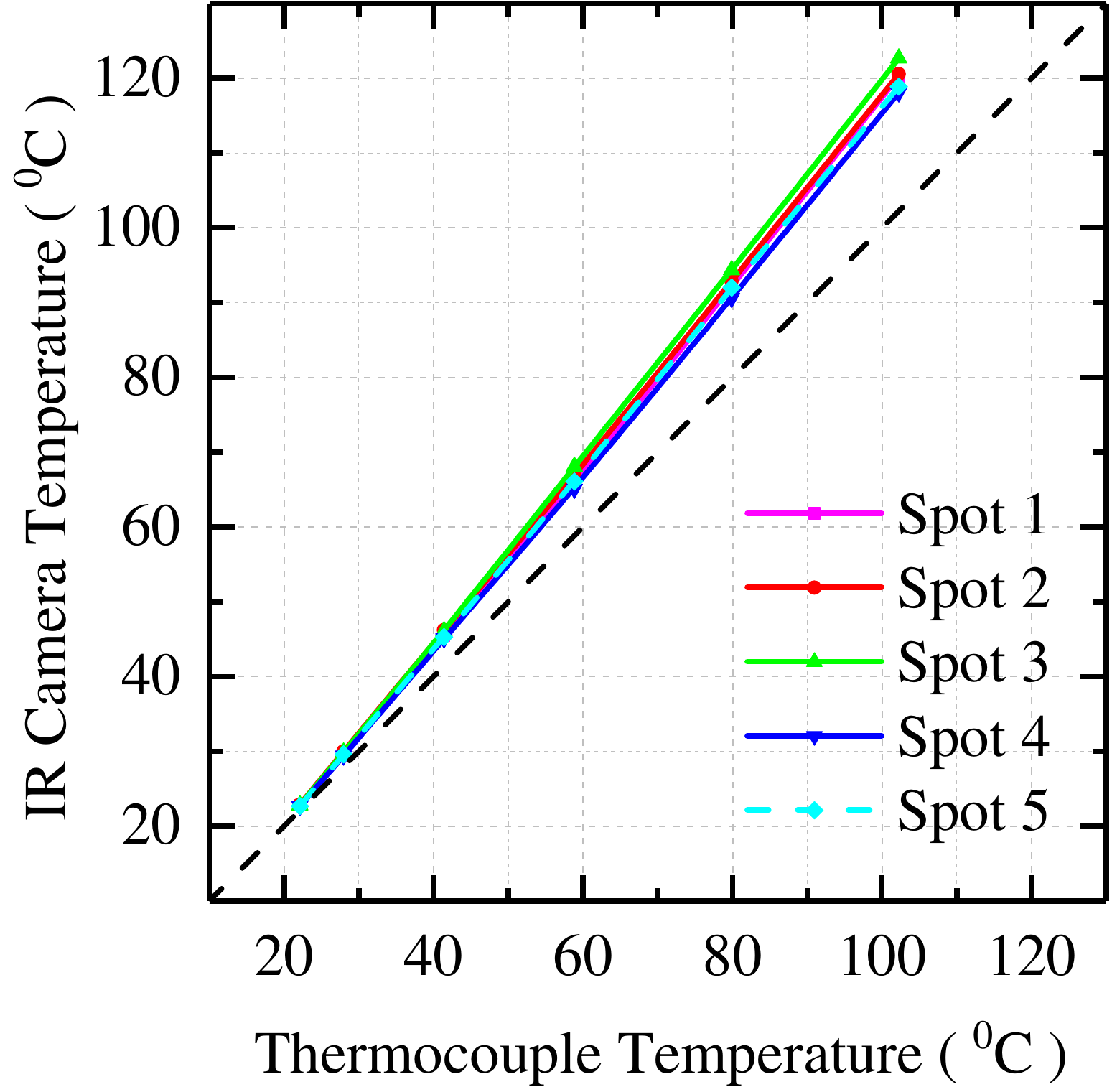}
\caption{Figure shows the steady-state relationship between the IR Camera recorded temperature and the base thermocouple temperature at different power settings for the five randomly chosen spots on the empty copper pan.}
\label{calibration}
\end{figure}

The strength of our work lies in our simplistic approach. In Figure~\ref{setup} we present a detailed outline of our experimental setup. A thin layer of Silicone oil is heated in a copper pan~\footnote{For more information about the properties of the sample fluid please refer~\cite{2018arXiv181206002C,shinetsu}}. A J-type thermocouple (T1) attached to the bottom of the copper pan measures its base temperature, $T_{bottom}$. Copper being a good conductor of heat allows the flow of energy through it with minimal thermal resistance. The average diameter of the pan is $0.225~m$. The pan is heated from the bottom by an electric resistance heater ($37.5\pm 0.5~\Omega$). The top cover is made up of wood and has inlet and outlet ducts for forced convective heat transfer. The two thermocouples T2 and T3 measure the temperature of the incoming and outgoing gas respectively. The bottom rest, also made up of wood has a cavity with a recess on which the copper pan sits snugly. The wooden base rests on top of a block of polyurethane foam. An IR camera, placed concentrically above the copper pan captures the real-time thermal images from a height ($\geq 0.7~m$). Each thermal image has its own temperature scale. In order to calibrate the IR camera with the base thermocouple (T1), the empty copper pan is heated and the temperature of five randomly chosen points on the copper pan are recorded at different power settings of the resistance heater at steady-state. The thermocouple temperature recorded by T1 is then compared with the IR camera recorded temperature for the five spots. The thermocouple data can be viewed in the NI Signal Express software while the IR camera data is viewed in the FLIR Software. The FLIR software allows remote accessibility like, real-time display, region/point selection and spatial statistics. In Figure~\ref{calibration}, we present the calibration curves for the five spots comparing the IR camera-recorded temperature with the base thermocouple temperature. The calibration error thus estimated is used to adjust the IR camera temperature scale in accordance with the base thermocouple.

Once the IR camera is calibrated one can set up the system to perform a series of experiments to gather thermal data. A high viscosity silicone oil is used for the purpose of the current study~\cite{2018arXiv181206002C,shinetsu}. The fluid is placed in the copper pan and it is then heated by regulating the externally applied voltage. It is always ensured that a large pan diameter ($2R$) to fluid-film thickness ($l_z$) is maintained, $2R/l_z\simeq~225~mm/5~mm\sim45$, as the goal
is to have convection cells over as wide as an area possible for the thermal imaging to yield significant temperature statistics. It is also ensured that the applied power never exceeds a value that can change the molecular configuration of the oil through burning or render the system into a turbulent regime through boiling. Since, our study is focused within the non-turbulent regime, the Rayleigh Number, $Ra < 3500$. For more details about the operational regime of the $Ra$ for the current scope of study please refer~\cite{2018arXiv181206002C}.

\section{Image Analysis}
An IR Camera detects heat emitted from an object and converts it into bits. Thus, every pixel has an allocated bit value between $0$ and $255$. The bit value determines the intensity of the pixel with $0$ being the `coldest' pixel in the image and $255$ being the most intense or the `hottest'. The FLIR T62101, used in this study has a resolution of $320\times 240$ pixels ($=76800$ pixels) with a sensitivity less than $0.045^\circ$C. Therefore, every image ($I$) is a 2D array ($M\times N$) of $320\times 240$ elements with $256$ bit values distributed in between them. In this study, the IR camera is used to film the thermal profile of the top layer of the fluid-film. Using the IR camera temperature scale and the estimated calibration error, each of these bit values can be converted to respective temperature values through a linear interpolation as shown in~\ref{eqn1}.

\begin{equation}
\text{Temp ($^\circ C$)} = \frac{\text{Max Temp ($^\circ C$) - Min Temp ($^\circ C$)}}{255 - 0}\times\text{bit}~+~\text{Min Temp ($^\circ C$)} 
\label{eqn1}
\end{equation}

Each entry in the $M\times N$ matrix can now be denoted by a temperature $T_{ij}$ with the pair $(i,j)$ defining a particular pixel location on the thermal image. 

\section{Results and Discussion}
\begin{figure}[]
\centering
\includegraphics[scale=0.48]{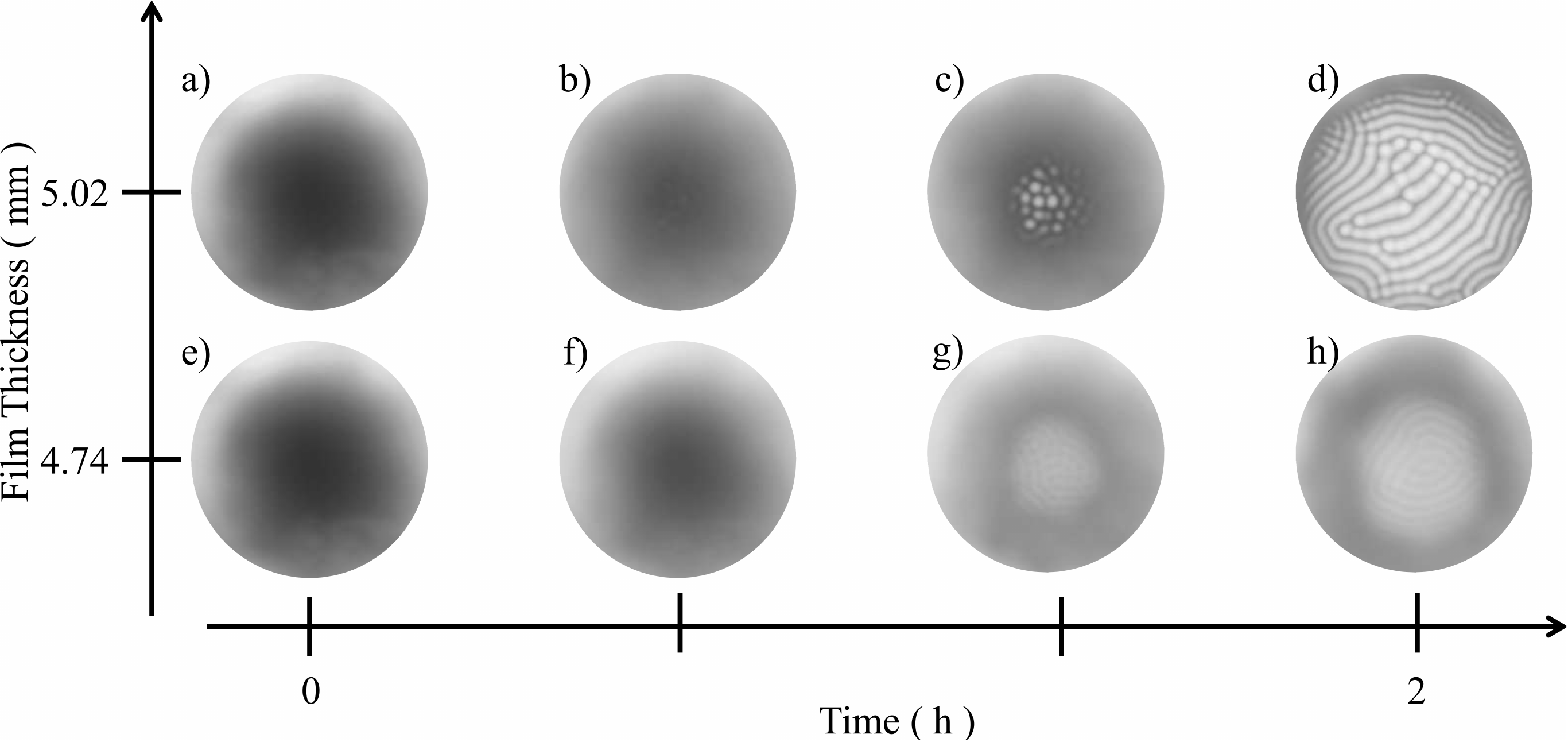}
\caption{Figure illustrates the temporal evolution of the Rayleigh-B{\'e}nard system from room temperature equilibrium to an out-of-equilibrium steady-state after two hours of constant heating at $66~W$ for two thicknesses of the fluid-film, $l_z = 4.74~mm$ and $5.02~mm$.}
\label{benard-images}
\end{figure}

This paper aims to illustrate the experimental and statistical methodologies to quantify spatio-temporal thermal fluctuations in a driven out-of-equilibrium steady-state system. A steady-state is achieved at a particular power by letting the fluid sample to gradually evolve from a room temperature equilibrium to an out-of-equilibrium state after two hours of constant heating. The snapshots of the thermal images are obtained using the IR camera for every $15^{th}$ of a second thus yielding $481$ images over the total duration of two hours. In Figure~\ref{benard-images}, we present a sample of the snapshots of the Rayleigh-B{\'e}nard system at different instances in time for the two film-thicknesses at a fixed power of $66~W$.

\begin{figure}[]
\centering
\includegraphics[scale=0.4]{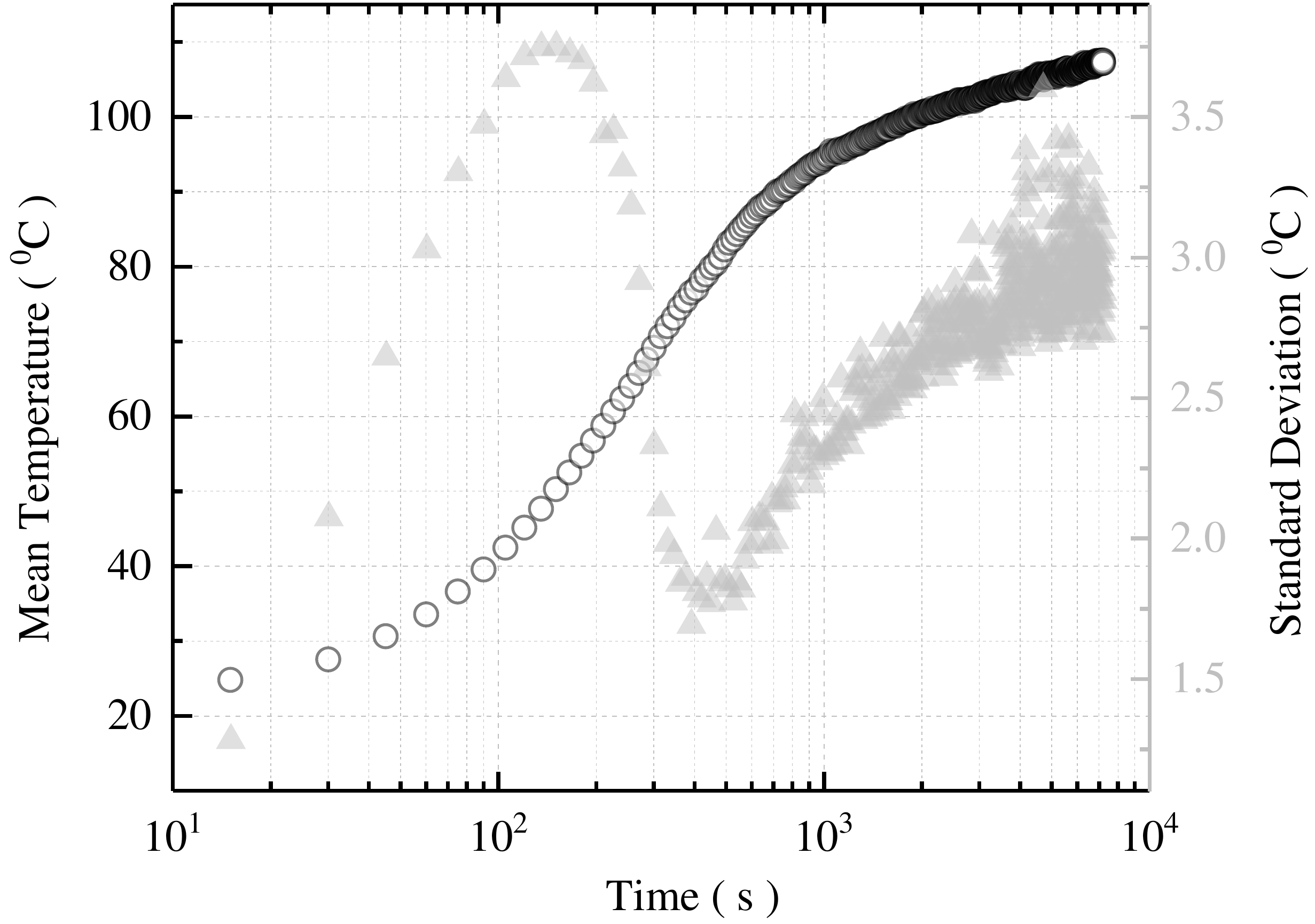}
\caption{Figure illustrates the mean and standard deviation of the temperature from the temperature matrix as a function of time as the the Rayleigh-B{\'e}nard system evolves from room temperature equilibrium to an out-of-equilibrium steady-state after two hours.}
\label{example-stat}
\end{figure}

\begin{figure}[]
\centering
\includegraphics[scale=0.6]{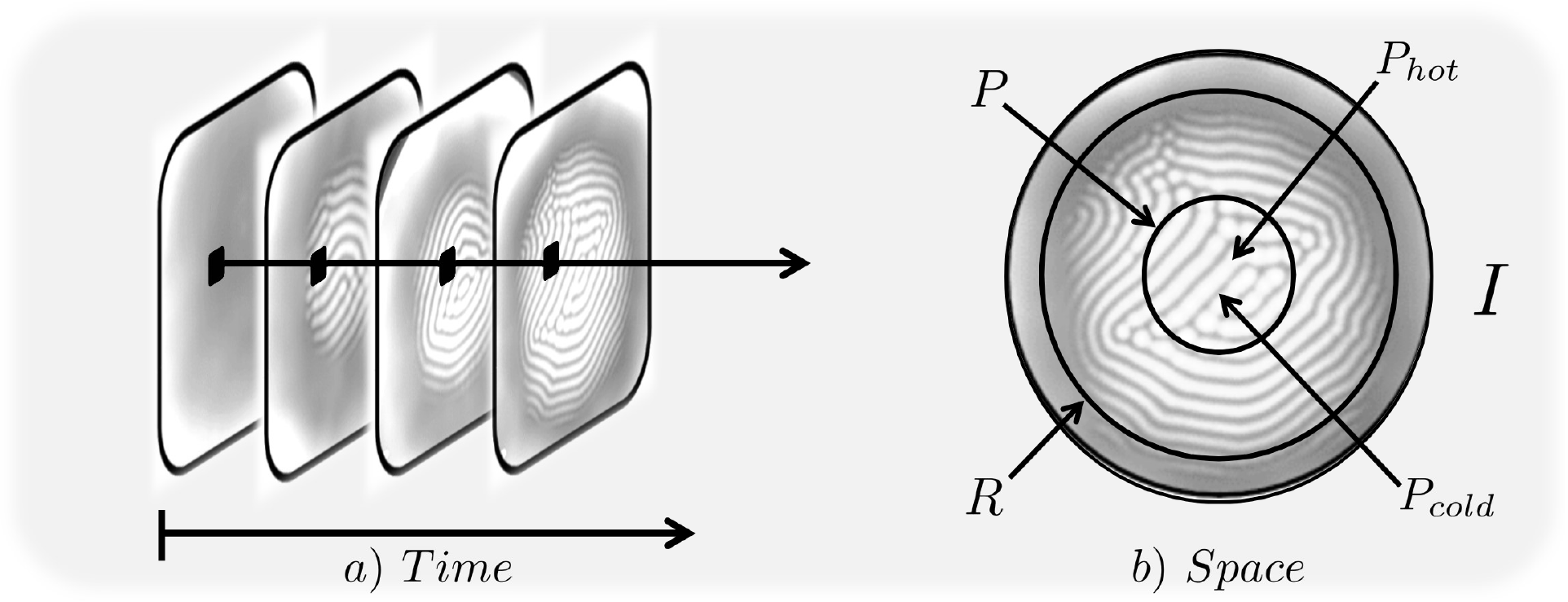}
\caption{a) Figure illustrates the temporal analysis strategy of a region of interest as a function time as the Rayleigh-B{\'e}nard system evolves from room temperature equilibrium to an out-of-equilibrium steady-state. b) Figure shows the regions of
interest for the spatial analysis on a steady-state image. The thermal image ($I$) is comprised of the annular region without any emergent patterns ($R$) and the pattern region ($P$). The upward (bright spots) and downward plumes (dark spots) inside $P$ are denoted by $P_{hot}$ and $P_{cold}$ respectively.}
\label{space-time}
\end{figure}

Each of these images are converted into a $M\times N$ temperature matrix on which statistical analyses are performed by choosing regions of interest. To select a square or a rectangular region, $(x_1, y_1)$ the coordinates of the upper left edge and $(x_2, y_2)$ the coordinates of the lower right edge are specified. To select a circular or an annular region, the center point and radii (internal and external) are specified with constraints on the spatial location of the $(x,y)$ coordinates in the region of interest. In summary, we can isolate regions of interest in each of the thermal images and thus perform statistical analysis spatially to obtain descriptive statistics such as, mean, standard deviation, kurtosis etc. In Figure~\ref{example-stat}, we illustrate as an example plot the mean and standard deviation of an arbitrary region of the top film temperature as a function of time as the the Rayleigh-B{\'e}nard system evolves from room temperature equilibrium to an out-of-equilibrium steady-state. To better illustrate the analysis strategy we present a pictorial representation in Figure~\ref{space-time}. In Figure~\ref{space-time}a, a specific region of interest is followed in time in order to extract temporal statistics whereas, in Figure~\ref{space-time}b, a region of interest is specified in a steady-state image in order to extract spatial statistics. The annular non-convective region is denoted by $R$, whereas the pattern region by $P$. The sub-regions within $P$ are denoted as $P_{hot}$ and $P_{cold}$ corresponding to updrafts (or upward plumes) and downdrafts (or downward plumes). For an arbitrary region on the top film the mean and standard deviation are calculated as, $\langle T\rangle = \frac{1}{N}\Sigma_{i,j}T_{ij}$ and $\sigma_T = \sqrt{\frac{\Sigma_{i,j}(T_{ij} - \langle T\rangle)^2}{N-1}}$.

\begin{figure}[]
\centering
\includegraphics[scale=0.45]{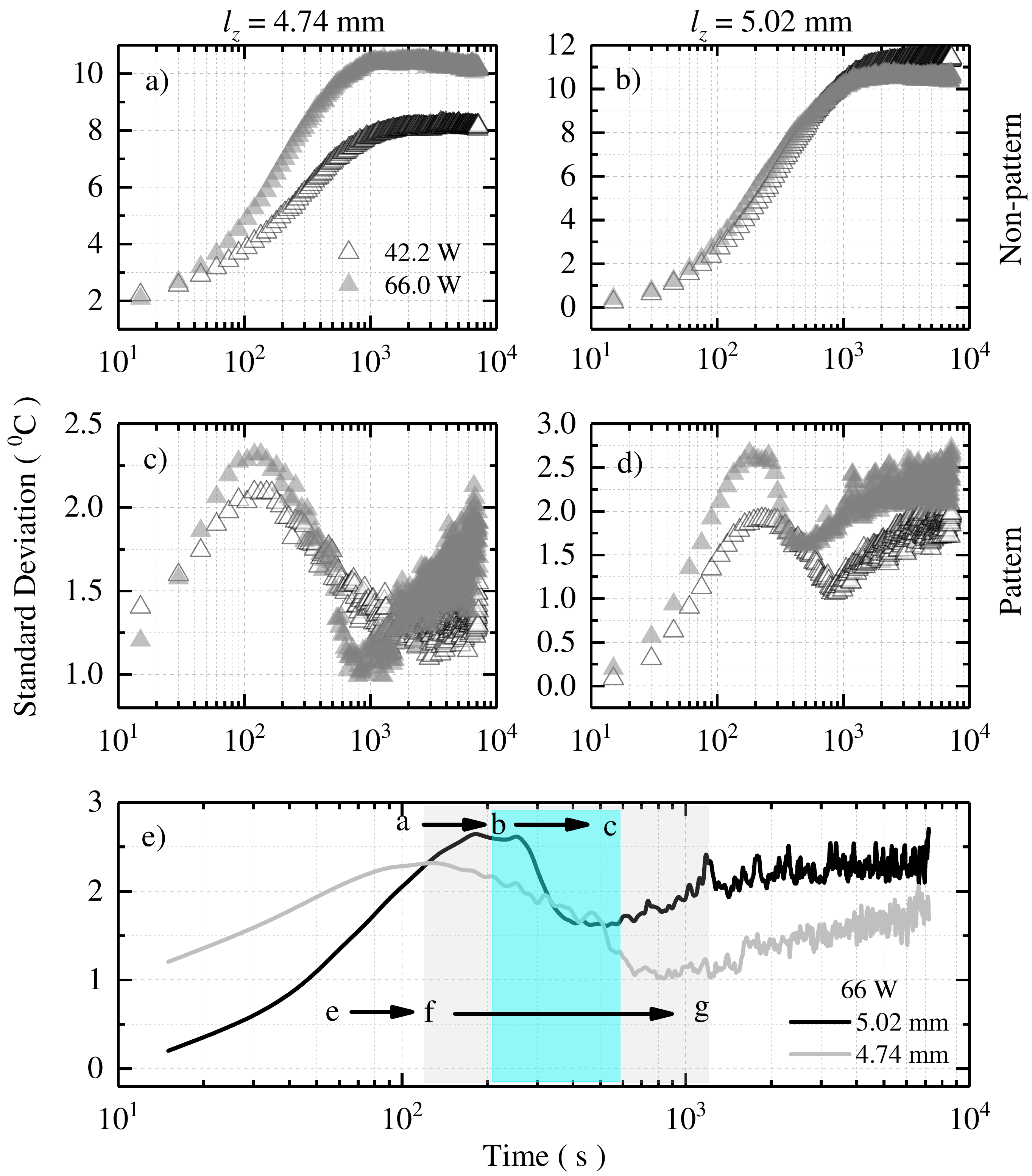}
\caption{a), b) Figure shows the functional relationship between the standard deviation of the temperature for the non-pattern region with time as the Rayleigh-B{\'e}nard system evolves from a room temperature equilibrium to an out-of-equilibrium steady-state for $l_z = 4.74~mm$ and $5.02~mm$ at $42.2~W$ and $66~W$. In c) and d) the data is shown for the pattern region. In e) a comparison is done between the two thickness at $66~W$ with the time windows identified: cyan for $5.02~mm$ and light grey for $4.74~mm$. The image label snapshots are from Figure~\ref{benard-images}.}
\label{std-dev}
\end{figure}

In Figure~\ref{std-dev}, we take our statistical analysis a step further by plotting the standard deviation of the temperature, a measure of the thermal fluctuation, as a function of time by isolating the pattern and non-pattern regions for the two thicknesses at $42.2~W$ and $66~W$. We can see contrasting trends between Figure~\ref{std-dev}a and~\ref{std-dev}b. In Figure~\ref{std-dev}a and~\ref{std-dev}b, the standard deviation increases with time till saturation as the system reaches a steady-state. Whereas, in the case of the pattern region in Figure~\ref{std-dev}c and~\ref{std-dev}d, the standard deviation first shows a decline when the pattern starts to form but is not yet visible, followed by a dip at the point a stable visible pattern starts emerging. We further illustrate this in Figure~\ref{std-dev}e, where we compare the temporal evolution of the standard deviation of the temperature for the two thicknesses at $66~W$. The shaded boxes in the plot identify the time windows when the standard deviation starts to decline and then increasing again. With increasing power we can observe that the standard deviation of the temperature grows in time and the time window for the standard deviation shortens as $Ra\sim\Delta T$. For a thicker fluid film the time window is observed to be shorter as, $Ra\sim l^3$. Thus, a critical local Rayleigh Number is achieved faster. One can also identify the beginning and the ending of the time window with the thermal image snapshot labels from Figure~\ref{benard-images}. The window in cyan identifies the transition from $b\rightarrow c$ for $5.02~mm$, whereas the window in light grey identifies the transition from $f\rightarrow g$ for $4.74~mm$. The images $a$ and $e$ are equilibrium snapshots (at $t=0$) taken for both the thicknesses respectively. 

\begin{figure}[t]
\centering
\includegraphics[scale=0.5]{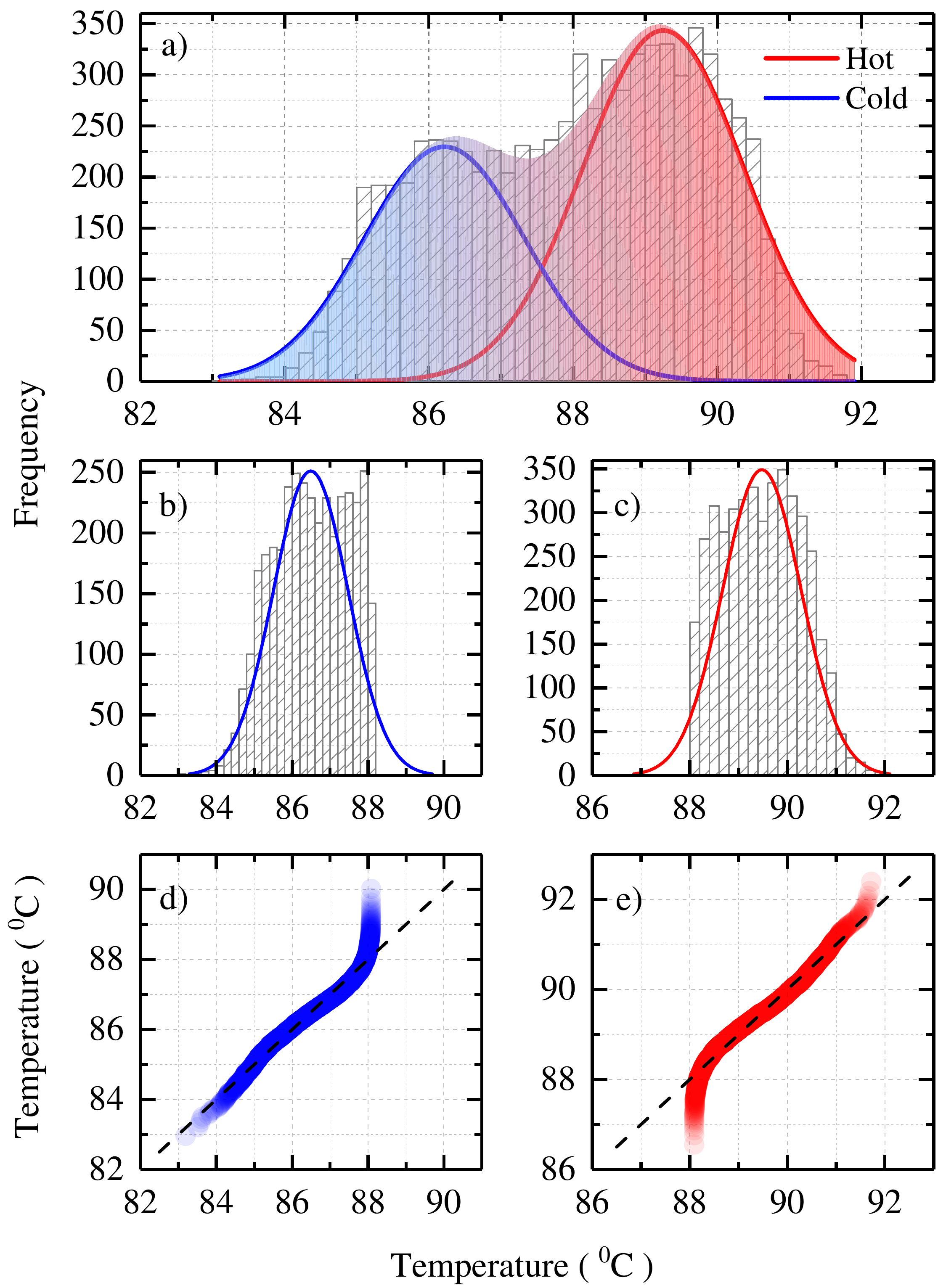}
\caption{a) Figure shows a bimodal distribution of the temperature frequency distribution over a region of pattern. Two independent Gaussian fits are performed which are identified as `hot' (in red) and `cold' (in blue). The shaded region is the cumulative function obtained from the two independent fits. In b) and c) a threshold temperature, $T_{th}$ is used to slice the data into two separate 1D arrays and their frequency histograms are plotted with the respective Gaussian fit functions. The normality is further tested in d) and e) by plotting the respective Q-Q plots at $95\%$ CI.}
\label{bimodal}
\end{figure}

We perform spatial statistical analysis on our steady-state images, i.e. for a particular thickness at a fixed power, we choose the last image (snapshot taken after two hours). We isolate pattern and non-pattern regions by defining circular and annular regions of interest on the thermal images. Once the regions are specified, a 1D array is created to store the temperature values, $T_{ij}$ at every pixel location. Once a 1D array of all temperature values, $T_{ij}$ is obtained a frequency histogram is generated to model the distribution of temperature spatially for our regions of interest. In Figure~\ref{bimodal}a, we show the histogram distribution when our region of interest falls over the patterns in the thermal images. A clear bimodality is observed in the frequency histogram. We identify the two peaks and perform two independent Gaussian fits, $\mathcal{N}(\mu_k , \sigma_k^2)$ on the data: $\mathcal{N}(89.25\pm 0.089, 1.25)$ for the `hot' region (in red) and $\mathcal{N}(86.21\pm 0.13, 1.24)$ for the `cold' region (in blue). The shaded region enclosing both the Gaussian fits is a cumulative fit function. We have shown in our previous works that a kernel density estimate provides a better description of the shape of the frequency data, $\hat{f}(T) = \frac{1}{nh}\Sigma_i K(\frac{T - T_{i}}{h})$. However, a kernel smoothening with Gaussian kernels ($K(\cdot)$) does not present any intuitive understanding of the physics beyond the shape of the estimate function. We identify the point where the two independent Gaussian fits intersect each other. We set the temperature at the point of intersection of the two curves as a threshold ($T_{th}$) to slice the frequency data into two regimes: `hot' and `cold'. Two independent 1D arrays are created and the temperature frequency data is binned accordingly: if $T_i\leq T_{th}$ then the list of all $T_i$ is in `cold' region and if $T_i > T_{th}$ then the list of all $T_i$ is in `hot' region. We look at their independent statistics in Figure~\ref{bimodal}b and~\ref{bimodal}c. The histograms are clearly normally distributed with $\mathcal{N}(89.48 , 0.64)$ for the isolated `hot' region and $\mathcal{N}(86.49 , 0.96)$ for the isolated `cold' region. One can see that the descriptive statistics of the isolated regions is well within one standard deviation of the continuous region with $95\%$ CI. To further elucidate the normal behavior we show the Q-Q plots for the isolated regions in Figure~\ref{bimodal}d and~\ref{bimodal}e. One can clearly note that at $T_i \sim T_{th}$ the curves depart from normality whereas everywhere else they are in agreement with the normal nature of the frequency distribution. We discuss this idea in our previous work as the coexistence of multiple equilibrium points in an out-of-equilibrium steady-state system~\cite{chatterjee2018many,2018arXiv181206002C}.



\begin{figure}[]
\centering
\includegraphics[scale=0.35]{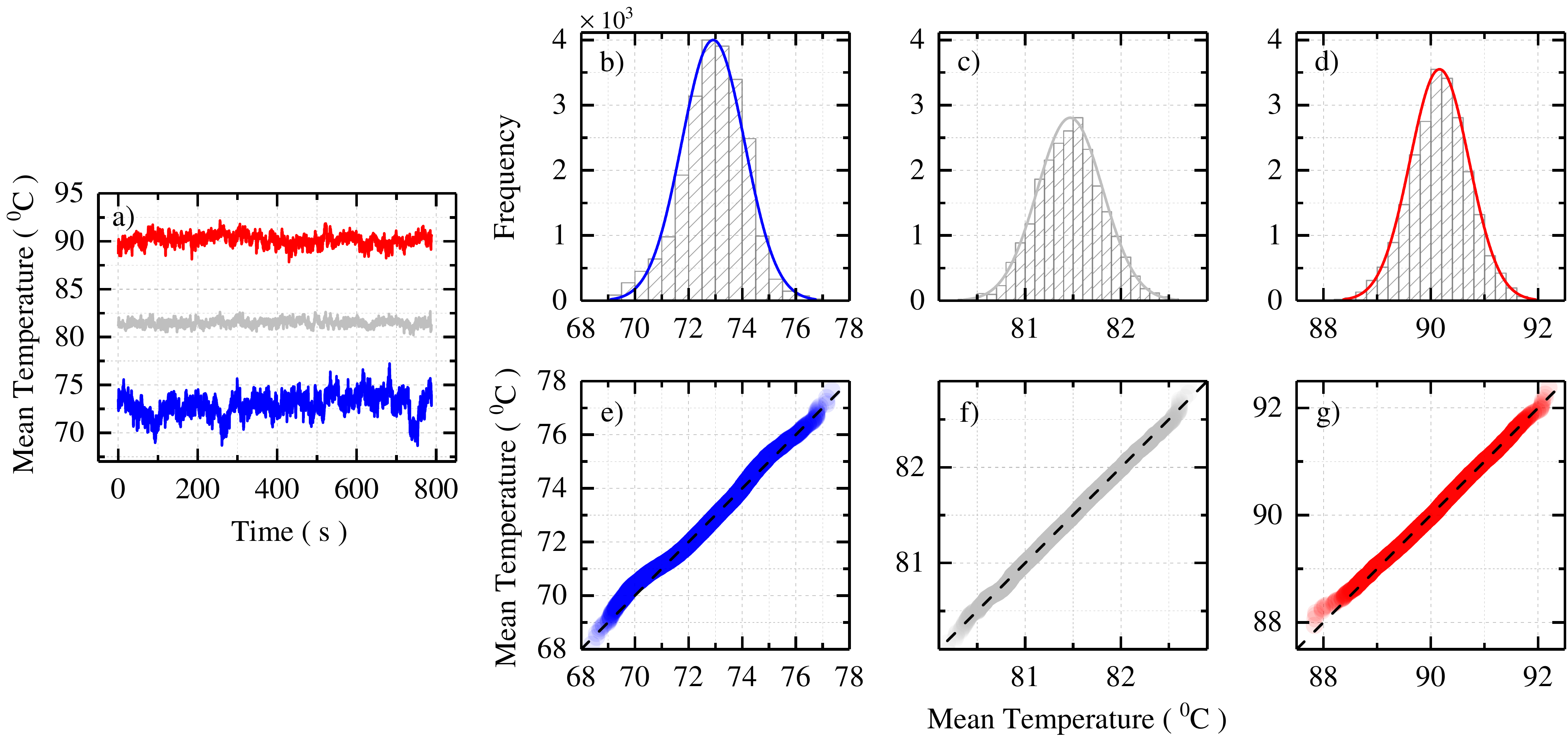}
\caption{a) Figure shows the spatially averaged temperature in different regions of interest as a function of time in a time-series plot. In b), c), and d) the time-series data is plotted as frequency histograms for `cold' (in blue), whole (in light grey), and `hot' (in red) regions respectively. In e), f), and g) the Q-Q plots are shown for each case to test for normality.}
\label{time-series}
\end{figure}

In our previous work we had claimed that the spatial symmetry is broken although temporal symmetry is preserved. In order to look at the temporal behavior of the Rayleigh-B{\'e}nard system at out-of-equilibrium steady-state, we let the system evolve beyond the steady-state ($\geq 2h$) and capture a real-time video. For the purpose of this study a $15$ minute video was recorded at $30$ frames/second thus yielding $27000$ frames. We show the time-series data in Figure~\ref{time-series}a. To obtain robust temporal statistics, regions of interest were selected on the thermal images and were spatially averaged across frames. In Figure~\ref{time-series}b,~\ref{time-series}c, and~\ref{time-series}d we plot the frequency histograms for the `hot' (in red), `cold' (in blue) and the whole region (in light grey). These histograms are clearly normally distributed: $\mathcal{N}(90.16\pm 0.0036 , 0.30)$ (red), $\mathcal{N}(72.91\pm 0.007 , 1.36)$ (blue) and $\mathcal{N}(81.47 , 0.11)$ all at $95\%$ CI. The statistical mean of the whole region is found to be very close to the average of the statistical means of the respective `hot' and the `cold' regions. To further ascertain the normality of the frequency histograms, Q-Q plots are shown for the each of the regions of interest in Figure~\ref{time-series}e,~\ref{time-series}f, and~\ref{time-series}g. 

\begin{figure}[]
\centering
\includegraphics[scale=0.5]{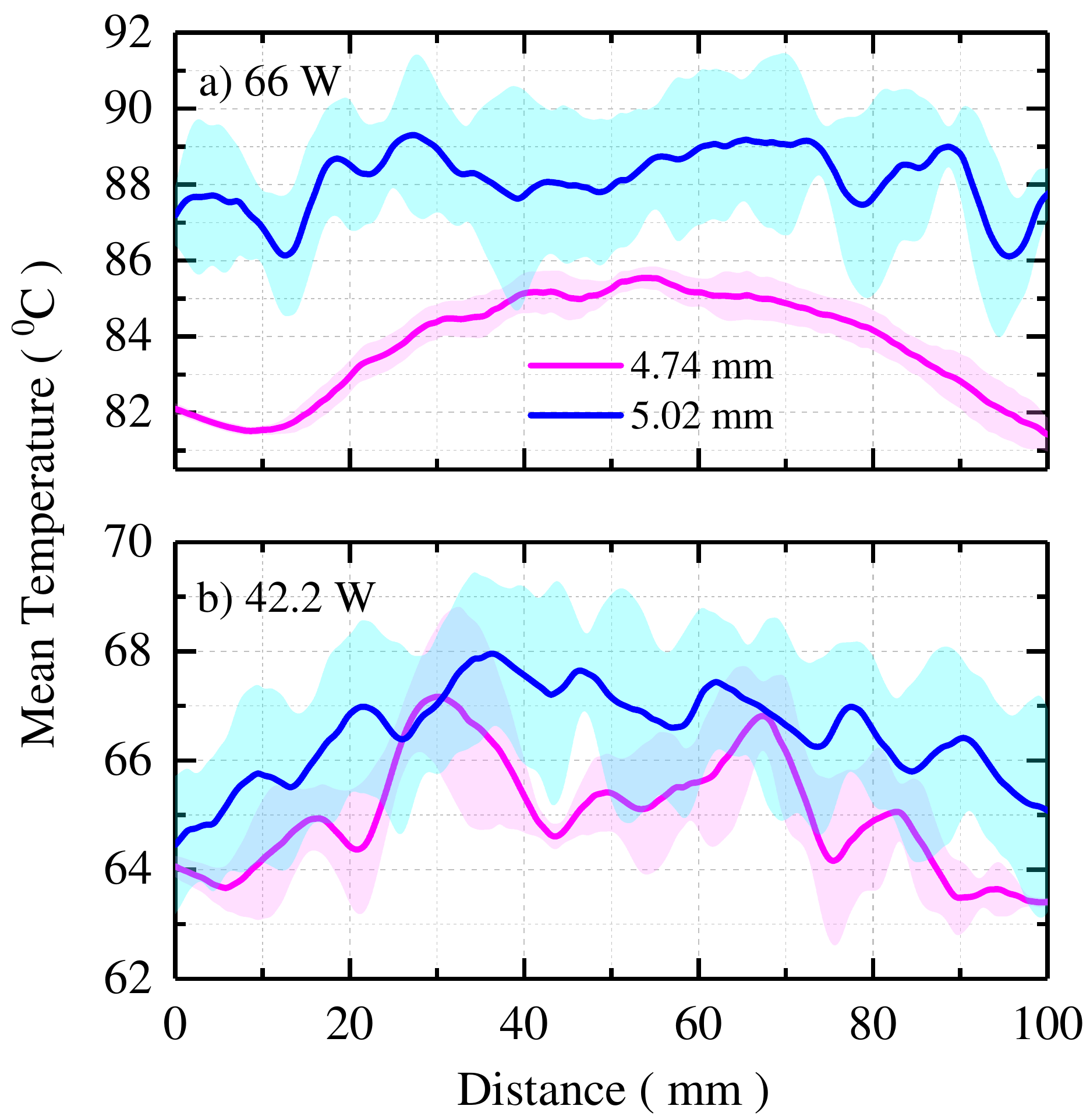}
\caption{a) Figure shows the mean thermal profiles of six spatially averaged horizontal lines for $L_z = 5.02~mm$ (blue) and $4.74~mm$ (magenta) at $66~W$. b) Figure shows the mean thermal profiles of six spatially averaged horizontal lines for $L_z = 5.02~mm$ (blue) and $4.74~mm$ (magenta) at $42.2~W$. The shaded bands about the mean thermal profiles represent the standard deviation.}
\label{linecut}
\end{figure}

It is clear by now that the thermal profile of the top layer of the fluid film is non-uniform. In order to visualize the modulation in the temperature as a function of distance several line cuts are performed on the thermal images. These line cuts are constructed from the image matrix by choosing 1D arrays of row/column data. In Figure~\ref{linecut}, we plot spatially averaged thermal profiles along six horizontal line cuts aligned parallel to each other. 

As expected, the mean temperature in Figure~\ref{linecut}a is greater than the mean temperature in~\ref{linecut}b. The modulation in the thermal profiles describe the thermal-field heterogeneity. The flatness of the thermal profile allows us to identify spatial correlation lengths of thermal fluctuations in the system. On comparing Figure~\ref{linecut}a and~\ref{linecut}b, it can be observed that at higher power ($66~W$) the thermal profiles tend to be more uniform over longer length-scales than at lower power ($42.2~W$). There are two reasons for this observation. One, at lower power the pattern occupies a smaller area about the center while the majority of the peripheral region being `cooler' thus implying more thermal heterogeneity as one approaches the periphery from the center, and two, at lower power the emergent patterns are not fully developed yet and hence randomly oriented. Thus, the modulation is uneven and hence chaotic. At higher power, the emergent patterns are fully developed and exhibit a closed packing thus being more homogeneous. 

In order to reconstruct the complete geometry of the thermal field we need to extract the thermal profile and integrate over the whole region. In order to execute this, thermal profiles of normalized integrated intensities around concentric circles are plotted as a function of distance from a point in the image. This point of reference is defined by the center of the rectangle that bounds the region of interest. The position of this point can be modified by user defined commands. The temperature at any given distance from this point represents the sum of the pixel values around a circle, whose radius is the distance from the point. The integrated intensity is then divided by the number of pixels in the circle thus yielding normalized density values. This profile is plotted as a function of distance by defining a starting and an integration angle to perform a radial averaging. The radial averaging allows us to identify the distance between subsequent maxima and minima on the thermal field. 

\begin{figure}[]
\centering
\includegraphics[scale=0.7]{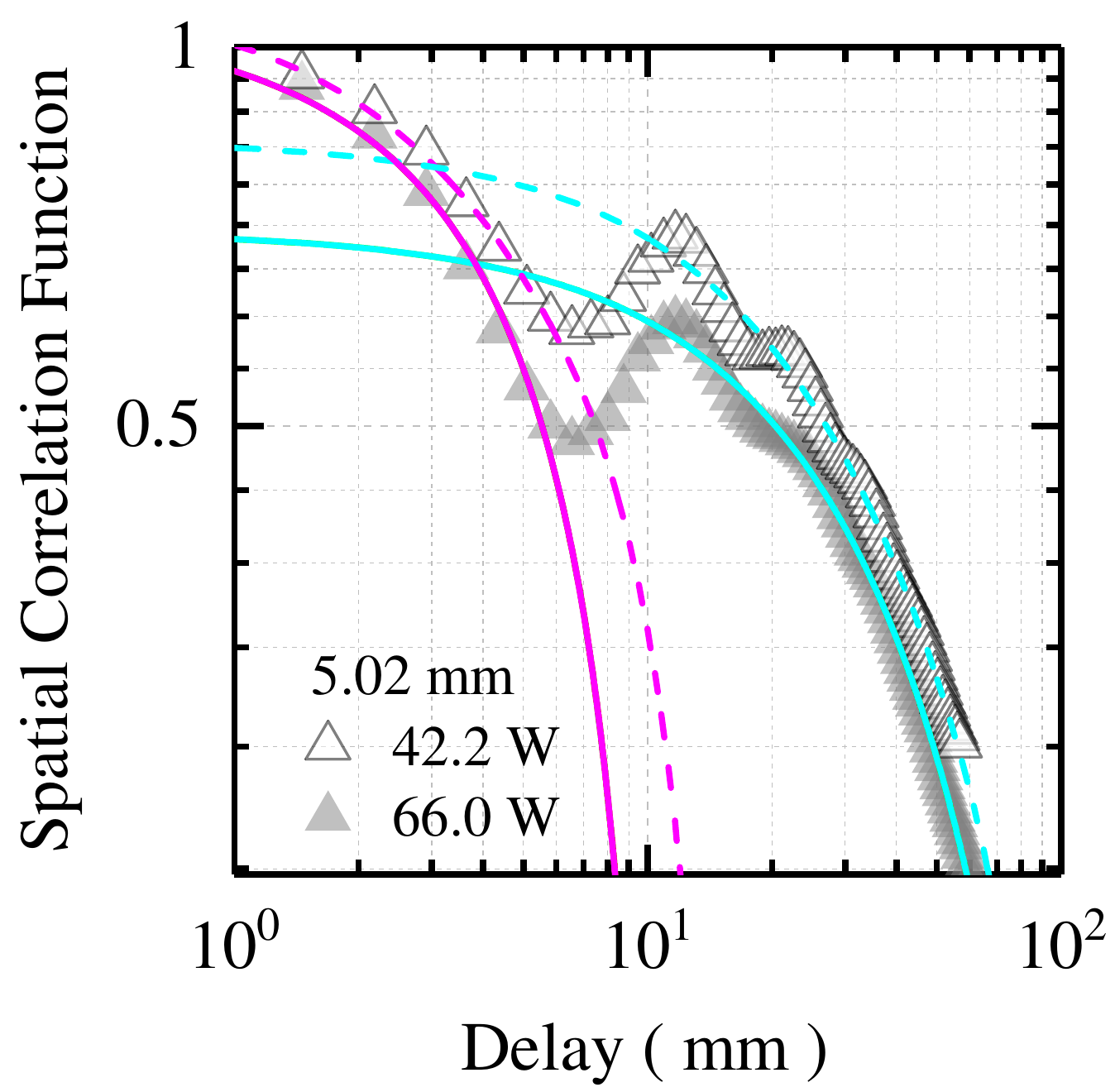}
\caption{Figure shows a typical spatial autocorrelation function at steady-state for the $5.02~mm$ sample at $42.2~W$ and $66~W$. The exponential fits (shown in cyan and magenta) are used to obtain the respective correlation lengths. Note that at higher power the correlation length decreases as patterns emerge.}
\label{acf}
\end{figure}

However, to extract meaningful length-scales from the entire thermal profile we need a more sophisticated technique. Therefore, we employ a spatial two-point autocorrelation to extract useful information about the emergent length scales from the system. It is done by comparing the frequencies of values in the temperature matrix, and then finding the most dominant frequencies. In this case, the autocorrelation function analyzes the 2D temperature profile matrix and finds correlations based on length, at an arbitrary delay, $r$ away from each element in the matrix. The program is written following the Weiner-Khinchin theorem which relates the autocorrelation function to the power spectral density via the Fourier transform~\cite{wiener1930generalized,mears2019stochastic}. The 2D spatial autocorrelation is executed using Python's numpy library. We make use of its 2D fast Fourier transform and its complex conjugate functions, $\texttt{fft.fft2}$ and $\texttt{conj}$. The matrix is normalized to a range of values $\{0,1\}$. Numpy provides another useful function called $\texttt{fft.fftshift}$ which moves the zero coordinate to the center of the domain. Implementing this outputs a much smoother autocorrelation function as output. In order to convert the 2D autocorrelation function into a 1D radius based autocorrelation, all the elements inside the circle of arbitrary radius $r$ around the center of the matrix are averaged and the normalized $\mathcal{G}_2(r)$ is plotted as a function of this delay in Figure~\ref{acf}. The spatial autocorrelation data is fitted with an exponential decay function, $\mathcal{G}_2(r)\sim\exp(-r/\xi)$ where $\xi$ is the correlation length. In Figure~\ref{acf}, we can see that there exists atleast two correlation lengths, $\xi\sim 4-6~mm$ and $\xi\sim 30-50~mm$. The larger correlation length captures coarser (more uniform) length scales whereas the smaller one captures finer length scales. One can compare Figure~\ref{acf} with Figure~\ref{linecut}a to check for consistency and refer to the discussion presented there. 

%
%
    %

\begin{figure}[]
\centering
\includegraphics[scale=0.5]{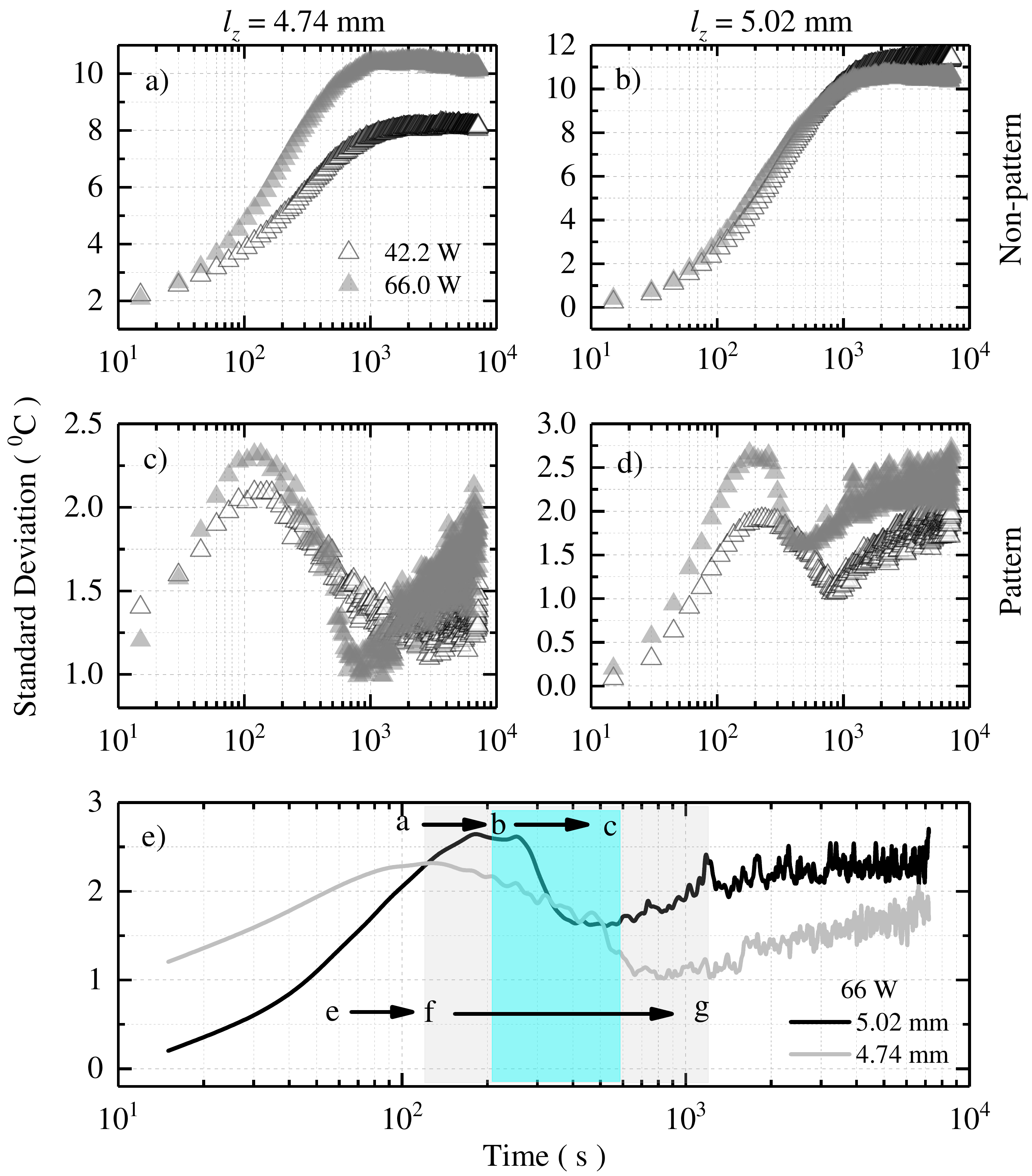}
\caption{a) Figure shows the spatial autocorrelation data along with exponential fits for $4.74~mm$ sample at steady-state. b) Figure shows the distribution of the two correlation lengths as the $4.74~mm$ sample evolves into an out-of-equilibrium steady-state from room temperature equilibrium. The two distribution functions: normal (in magenta) and lognormal (in red) are also shown in the figure.}
\label{corrlendist}
\end{figure}

We saw in our previous discussion that a steady-state thermal profile has multiple emergent length-scales. In particular, it was observed in Figure~\ref{acf} that there are atleast two dominant length-scales. When the Rayleigh-B{\'e}nard system is at room temperature equilibrium, it is thermodynamically homogeneous over large length-scales. Once, visible patterns emerge due to the system being driven out-of-equilibrium, finer length-scales start emerging on top of the intrinsic larger length-scale, as shown in Figure~\ref{acf} and~\ref{corrlendist}a. In our previous work we had discussed the connection between the emergence of the finer length-scales and the decline in the standard deviation of the temperature (see Figure~\ref{example-stat} and~\ref{std-dev})~\cite{2018arXiv181206002C}. A question that naturally follows is how these typical length-scales are distributed (in time) as the system evolves from a room temperature equilibrium to an out-of-equilibrium steady-state. The spatial autocorrelation script is implemented on all the $481$ temperature matrices (or snapshots of the thermal images) while simultaneously fitting the spatial correlation data with exponential fits. The first data point of the spatial correlation dataset is always equal to one hence this point acts as an anchor for the first exponential fit. In order to obtain the second characteristic length, a search algorithm is implemented on the spatial correlation datset to search for a local minima followed by a local maxima. The local maxima acts as the anchor for the second exponential fit. It is also made sure that the second exponential fit undercuts the first exponential fit. The correlation length from the first exponential fit gives an estimate of the finer length-scales whereas, the second exponential fit gives an estimate of the larger (equilibrium or close to equilibrium) length scales in the system. The frequency histograms are then plotted as shown in Figure~\ref{corrlendist}b. It is interesting to note that the larger length scales are lognormally distributed thus implying a positive skewness. Therefore, as the system ages (driven to a steady-state) finer length-scales emerge and a symmetrical normal distribution shifts towards the left. To put things into perspective of the system's physical dimension, the smallest length-scale is estimated to be never less than $20~mm$ and the largest length-scale is never greater than $120~mm$ which is slightly above the radius of the copper pan. 

\section{Conclusion}
In this paper, we discuss thoroughly the computational and the statistical techniques that were employed to perform a first principles thermodynamic study on the non-turbulent steady-state Rayleigh-B{\'e}nard convection when driven out-of-equilibrium. The two key questions that form the basis of our previous studies are: (i) how does an out-of-equilibrium steady-state differ from an equilibrium state and (ii) how do we explain the spontaneous emergence of stable structures and simultaneously interpret the physical notion of temperature when out-of-equilibrium. We discuss in the very beginning of the paper our simplistic approach by laying out the details of the experimental setup. Thereafter, we discuss the details of the computational and statistical techniques. For the purpose of our analysis several programming software were used at different stages of the work like, C, Python, Matlab, ImageJ etc. From a statistical point of view, thorough descriptive statistical computations were performed to generate confidence intervals, linear and non-linear curve fittings were performed to model the large amount of experimental data and test goodness of fits. Image analysis techniques, like smoothening, parsing, slicing and matrix manipulations were also performed. We believe that this paper will add to the strength of our previous work by making it accessible and reproducible at the same time. We are very optimistic that this paper will serve as a repository of techniques, both experimental and computational, in advancing our knowledge about far-from-equilibrium thermodynamics. 

\section{Acknowledgment}
The authors thank Prof. Germano Iannacchione in the Department of Physics at Worcester Polytechnic Institute for his critical feedback on the first draft of the paper. A.C. also expresses his gratitude towards the Department of Mathematical Sciences at Worcester Polytechnic Institute for arranging a meeting with Prof. Giovanni Gallavotti, which provided new insights about the theoretical aspects of the system concerning transport phenomenon and irreversibility.

\end{document}